\documentclass[fleqn,usenatbib]{mnras}

\usepackage{newtxtext,newtxmath}

\usepackage[T1]{fontenc}

\DeclareRobustCommand{\VAN}[3]{#2}
\let\VANthebibliography\thebibliography
\def\thebibliography{\DeclareRobustCommand{\VAN}[3]{##3}\VANthebibliography}

\usepackage{graphicx}	
\usepackage{amsmath}	
\usepackage{hyperref}
\usepackage{subcaption}

\usepackage{scalerel,tikz}
\usetikzlibrary{svg.path}
\definecolor{orcidlogocol}{HTML}{A6CE39}
\tikzset{orcidlogo/.pic={
 \fill[orcidlogocol] svg{M256,128c0,70.7-57.3,128-128,128C57.3,256,0,198.7,0,128C0,57.3,57.3,0,128,0C198.7,0,256,57.3,256,128z};
 \fill[white] svg{M86.3,186.2H70.9V79.1h15.4v48.4V186.2z}
 svg{M108.9,79.1h41.6c39.6,0,57,28.3,57,53.6c0,27.5-21.5,53.6-56.8,53.6h-41.8V79.1z M124.3,172.4h24.5c34.9,0,42.9-26.5,42.9-39.7c0-21.5-13.7-39.7-43.7-39.7h-23.7V172.4z}
 svg{M88.7,56.8c0,5.5-4.5,10.1-10.1,10.1c-5.6,0-10.1-4.6-10.1-10.1c0-5.6,4.5-10.1,10.1-10.1C84.2,46.7,88.7,51.3,88.7,56.8z};
}}
\newcommand\orcidicon[1]{\href{https://orcid.org/#1}{\mbox{\scalerel*{
\begin{tikzpicture}[yscale=-1,transform shape]
\pic{orcidlogo};
\end{tikzpicture}
}{|}}}}

\newcommand{\aref}[1]{\hyperref[#1]{Appendix~\ref{#1}}}

\title{High-precision star formation efficiency measurements in nearby clouds}

\author[Hu et al.]{
Zipeng Hu,$^{\orcidicon{0000-0002-3758-552X}\,1}$\thanks{zphu.charles@gmail.com (ZPH)}
Mark R.~Krumholz,$^{\orcidicon{0000-0003-3893-854X}\,1,2}$
Riwaj Pokhrel $^{\orcidicon{0000-0002-0557-7349}\,3}$
and Robert A. Gutermuth $^{\orcidicon{0000-0002-6447-899X}\,4}$
\\
$^{1}$Research School of Astronomy and Astrophysics, Australian National University, Canberra, ACT 2611, Australia\\
$^{2}$ARC Centre of Excellence for Astronomy in Three Dimensions (ASTRO-3D), Canberra, ACT 2611, Australia\\
$^{3}$Ritter Astrophysical Research Center, Department of Physics and Astronomy, University of Toledo, Toledo, OH 43606, USA\\
$^{4}$Department of Astronomy, University of Massachusetts, 710 North Pleasant Street, Amherst, MA 01003, USA\\
}

\date{Accepted XXX. Received YYY; in original form ZZZ}

\pubyear{2021}

\begin{document}
\label{firstpage}
\pagerange{\pageref{firstpage}--\pageref{lastpage}}
\maketitle

\begin{abstract}
On average molecular clouds convert only a small fraction $\epsilon_{\rm ff}$ of their mass into stars per free-fall time, but differing star formation theories make contrasting claims for how this low mean efficiency is achieved. To test these theories, we need precise measurements of both the mean value and the scatter of $\epsilon_{\rm ff}$, but high-precision measurements have been difficult because they require determining cloud volume densities, from which we can calculate free-fall times. Until recently, most density estimates assume clouds as uniform spheres, while their real structures are often filamentary and highly non-uniform, yielding systematic errors in $\epsilon_{\rm ff}$ estimates and smearing real cloud-to-cloud variations. We recently developed a theoretical model to reduce this error by using column density distributions in clouds to produce more accurate volume density estimates. In this letter, we apply this model to recent observations of 12 nearby molecular clouds. Compared to earlier analyses, our method reduces the typical dispersion of $\epsilon_{\rm ff}$ within individual clouds from 0.35 dex to 0.31 dex, and decreases the median value of $\epsilon_{\rm ff}$ over all clouds from $\approx 0.02$ to $\approx 0.01$. However, we find no significant change in the $\approx 0.2$ dex cloud-to-cloud dispersion of $\epsilon_{\rm ff}$, suggesting the measured dispersions reflect real structural differences between clouds.
\end{abstract}




\section{Introduction}
\label{sec:intro}
Star formation is inefficient. A star forming region with little resistance to self-gravity will convert its gaseous mass to stars within a single free-fall time $t_{\rm ff} =  \sqrt{3\pi/32G\rho}$, where $\rho$ is the volume density; the ratio of the actual star formation rate to this maximum is called the efficiency per free-fall time $\epsilon_{\rm ff}$ \citep{Krumholz_2005}.
\citet{Zuckerman_1974} were the first point out that the observed star formation rate of the Milky Way as a whole implies that, averaged over all molecular clouds, $\epsilon_{\rm ff} \sim 0.01$. \citet{Krumholz_2007} extended this conclusion to the denser parts of molecular clouds traced by HCN, and more recent work has obtained similar means both for clouds traced by dust \citep[e.g.,][]{Heyer_2016} and for $\sim 100$ pc-scale patches in nearby galactic discs \citep[e.g.,][]{Utomo_2018}. The study-to-study dispersion in $\epsilon_{\rm ff}$ is $\approx 0.3$~dex, while the dispersion within any single study is about 0.3 -- 0.5 dex \citep{Krumholz_2019}.

Theoretical efforts to explain the origin of observed low $\epsilon_{\rm ff}$ values can be categorized into two main types. Some theories focus on galactic scale physical processes \citep[e.g.][]{Kim_2011, Ostriker_2011, Faucher_Giguere_2013}, while others are developed from internal star formation regulation processes within individual molecular clouds \citep[e.g.][]{Elmegreen_1994,Krumholz_2011,Federrath_2012}. Despite predicting similarly-low mean $\epsilon_{\rm ff}$ values, the two types of models yield significantly different estimates for the \textit{dispersion} of $\epsilon_{\rm ff}$ -- models regulated on the cloud scale generally predict much smaller dispersions than those regulated on the galactic scale \citep{Lee_2016, Krumholz_2020}. Therefore, it is important to measure cloud-scale $\epsilon_{\rm ff}$ values with enough fidelity to extract both its mean value and dispersion. The most accurate measurements to date are those of \citet{Pokhrel_2021}, who determine a median value $\log \epsilon_{\rm ff} = -1.59$ with a dispersion of 0.18 dex in a sample of 12 nearby molecular clouds.

However, most previous $\epsilon_{\rm ff}$ measurements, including those of \citet{Pokhrel_2021}, incur a substantial error when calculating the volume density, which is required for the free-fall time. The fundamental challenge is that the volume density is a three-dimensional quantity, which is not directly accessible in a two-dimensional observation. The most common practice in literature is to estimate the density by assuming that the area of interest is the projection of a uniform sphere, whose radius is equal to the mean radius of the projected shape. For a cloud with a total projected area $A$ and a total mass $M$, this approximation gives a density estimate $\rho_{\rm sph} = 3M/4\sqrt{A^3/\pi}$. This method has been used by a number of authors in the Milky Way \citep[e.g.,][]{Krumholz_2011b, Lada_2013, Evans_2014, Pokhrel_2021} and in the Large Magellanic Cloud \citep{Ochsendorf_2017}. While simple, this procedure likely introduces significant systematic errors, coming from two primary sources. One is that in the past decades, it has become clear that the interstellar medium is characterised mainly by filamentary structures \citep[e.g.][]{Schneider_1979, Dobashi_2005,Arzoumanian_2011, Andre_2014, Kainulainen_2016}, which results in elongated contours identified from column density maps. The mean density of such structure is likely to be different from that calculated with spherical assumption. Second, the free-fall time computed from the mean volume density is not identical to the mean free-fall time, because the relationship between the two, $t_{\rm ff} \propto \rho^{-1/2}$, is non-linear. The quantity of interest for $\epsilon_{\rm ff}$ is the mass-weighted mean of $1/t_{\rm ff}$, and this mean can differ substantially from the overall mean density \citep[e.g.][]{Hennebelle_2011,Federrath_2012,Federrath_2013,Salim_2015}.

\citet{Hu_2021} recently proposed an alternative approach to estimate the free-fall time weighted mean density $\rho_{\rm eff}$, which immediately yields the correct free-fall time for the purposes of estimating $\epsilon_{\rm ff}$. \citeauthor{Hu_2021} analyse star formation simulations from \citet{Cunningham_2018}, and show that one can estimate $\rho_{\rm eff}$ with higher accuracy than is obtained from the simple spherical assumption by making use of the full two-dimensional (2D) column density distribution, rather than simply its mean. Applied to simulations, their method corrects a $\approx 0.13$ dex overestimate and removes a $\sim 0.25$ dex scatter in $\epsilon_{\rm ff}$ caused by spherical assumption. In this paper, we apply this model to the observations of \citet{Pokhrel_2020, Pokhrel_2021} in order to derive higher-accuracy $\epsilon_{\rm ff}$ measurements than have previously been possible. We summarise the observations in \autoref{sec:observations}, describe how we apply the \citeauthor{Hu_2021} method to them in \autoref{sec:methods}, present the results of the analysis in \autoref{sec:results}, and draw conclusions in \autoref{sec:conclusions}.

\section{Observations}
\label{sec:observations}
Our data reduction and analysis method is described in  \citet{Pokhrel_2020}, and full details are provided there. Here we simply summarise for convenience. This study analyses 12 nearby star forming regions: Ophiuchus, Perseus, Orion-A, Orion-B, Aquila-North, Aquila-South, NGC 2264, S140, AFGL 490, Cep OB3, Mon R2, and
Cygnus-X. For each region we have a matched protostellar catalogue and cloud column density map. The latter are derived from \textit{Herschel}/PACS and \textit{Herschel}/SPIRE imaging at 160, 250, 350, and 500 $\mu$m, convolved to a common resolution \citep{Andre_2010}. To obtain the column density in each pixel, we fit the spectral energy distribution with a dust emission model where the only two free parameters are the temperature and the column density. The best-fitting column density can be equivalently expressed in column of $\rm H_2$ molecules, $N(\rm H_2)$, or column of gas mass $\sigma_{\rm gas}$, which are related by $\Sigma_{\rm gas} = 2m_{\rm H}/{X}N(\rm H_2)$, where $m_{\rm H}$ = $1.67 \times 10^{24}$ g is the hydrogen atom mass and $X = 0.71$ is the hydrogen mass fraction of the local interstellar medium \citep{Nieva_2012}. On the obtained column density map, we first mask pixels with the best-fitting dust temperature exceeds the Rayleigh–Jeans (R-J) limit, since such pixels are in the R-J limit for all the \textit{Herschel} bands, and thus the fits are highly uncertain. The exact temperature limits are provided in Table 2 of \citet{Pokhrel_2020}. Second, we mask pixels with derived column densities $N(\rm H_2) > 10^{23} \: \rm cm^{-2}$, because for these dust optical depth effects can be significant and thus fitted $N(\rm H_2)$ values may only represent lower limits. The effect of both masks is negligible in our results, since the number of masked pixels is very small, and, for many clouds, zero. 

The SESNA catalog (R.~Gutermuth et al., in preparation) we use for protostars is a combination of \textit{Spitzer} and Two Micron All-Sky Survey observations (2MASS; \citealt{Skrutskie_2006}), spanning about 90 $\rm deg^2$. For the farthest target Cygnus-X, the deeper UKIDSS \citep{Lawrence_2007} near-IR Galactic Place Survey \citep{Lucas_2008} is used. We mask parts of column density maps outside SESNA coverage. After subtracting field stars, sources with excess IR emission are classified into different types of young stellar object (YSO) and contaminant by using flux selections and a series of reddening-safe color \citep{Gutermuth_2009}. Hence the total number of protostars $N_{\rm PS}$ in any enclosed contour on the column density map can be calculated after contamination correction, whose detailed procedure is provided in \citet{Pokhrel_2020}.

\section{Methods}
\label{sec:methods}
\begin{figure}
    \includegraphics[width = \linewidth]{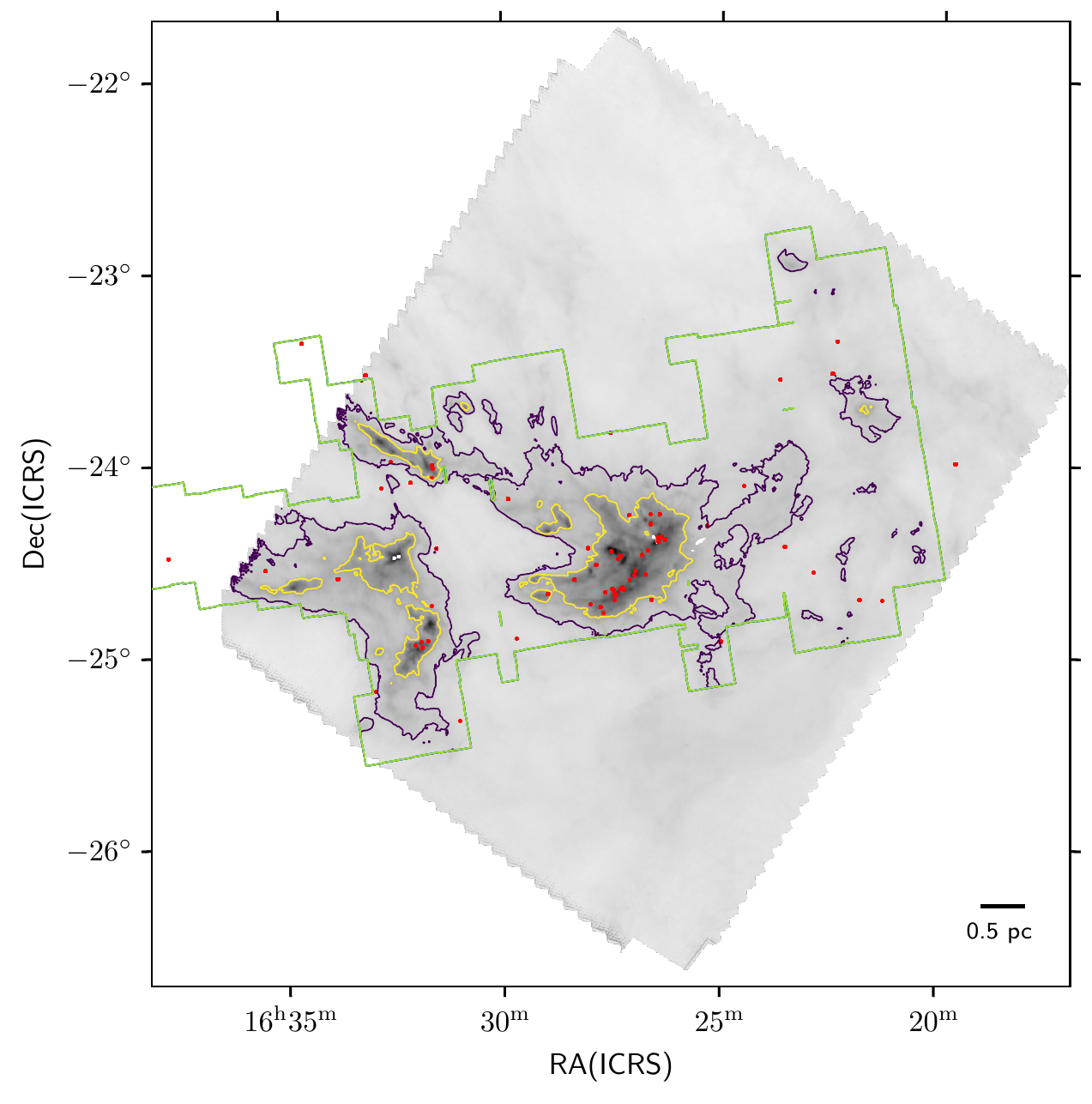}
\caption{A column density map of the Ophiuchus Cloud derived from \textit{Herschel} observations. The green contour is the Spitzer coverage area, and we we illustrate the largest contour that fits within this footprint in purple; the yellow contour is set at a level that encloses half as much mass as the purple one. The red dots are the positions of YSOs. 
} 
\label{fig: oph}
\end{figure}

The first step of our analysis is to generate column density contours across the full range of $N(\rm H_2)$ covered by the observations. For each of the 12 regions, we start by finding the lowest $N(\rm H_2)$ value such that all pixels with column density above it are inside the SESNA coverage; equivalently, we set the minimum value of $N(\rm H_2)$ to be the lowest possible choice such that the contour sits entirely within the SENSA footprint. We define $M_0$ as the total gas mass enclosed by this contour. We then draw 100 $N(\rm H_2)$ contours at higher $N(\rm H_2)$, with the levels chosen such that the total mass above each level is equally spaced from $M_0$ to $M_0/100$ with a step of $M_0/100$. We show an example of Ophiuchus cloud column density map with $N(\rm H_2)$ contours, protostar positions, and SESNA coverage area in \autoref{fig: oph}.

For every contour, we measure four values: total gas mass $M_{\rm gas}$, total area $A$, completeness-corrected total protostar number $N_{\rm PS}$ and the Gini coefficient $g$ of the surface density of all pixels within the contour (see \citealt{Hu_2021} for details). From these parameters we derive 4 additional quantities: mean gas surface density $\Sigma_{\rm gas} = M_{\rm gas}/A$, star formation surface density $\Sigma_{\rm SFR}$, the free-fall time derived from the spherical assumption $t_{\rm ff, sph}$ (see \autoref{sec:intro}), and the corresponding star formation efficiency per free fall time $\epsilon_{\rm ff,sph}$. To calculate $\Sigma_{\rm SFR}$, we assume the mean mass of protostars in SESNA to be $M_{\rm PS} \approx 0.5 M_{\rm \odot}$ \citep{Evans_2009}, and the mean duration of protostellar phase included in SESNA observations to be $t_{\rm PS} \approx 0.5$ Myr \citep{Dunham_2014,Dunham_2015}. Thus, $\Sigma_{\rm SFR} = N_{\rm PS}M_{\rm PS}/At_{\rm PS}$, and $\epsilon_{\rm ff, sph} = \Sigma_{\rm SFR}/(\Sigma_{\rm gas}/t_{\rm ff, sph})$.

Our next step is to apply the correction obtained by \citet{Hu_2021}, who find a relation between free-fall time weighted mean density $\rho_g$ and $\rho_{\rm sph}$:
\begin{equation}
    \rho_g = 10^{kg - b}\rho_{\rm sph},
    \label{eq: rho_g}
\end{equation}
with $k = 4.6$ and $b = 0.93$. Replacing $\rho_{\rm sph}$ with $\rho_{g}$ gives us the corrected values of free-fall time $t_{\rm ff, g}$ and star formation efficiency per free-fall time $\epsilon_{\rm ff, g}$. In order to accomplish our goal of obtaining high-accuracy estimates of $\epsilon_{\rm ff}$, it is important to understand the uncertainties in this relation. Although the values of $k$ and $b$ are fitted from exact simulation data without error bars, we can obtain estimate of the errors on these quantities from bootstrapping. We randomly choose half of the simulation sample to fit \autoref{eq: rho_g}, repeating this process $10^4$ times. We take our uncertainties to be the 16th and 84th percentiles of the fitted $k$ and $b$ values, which give $k = 4.6 \pm 0.1$ and $b = 0.93 \pm 0.03$. We show below that these uncertainties are small enough that they contribute negligibly to the overall error budget.

\begin{figure}
    \includegraphics[width = \linewidth]{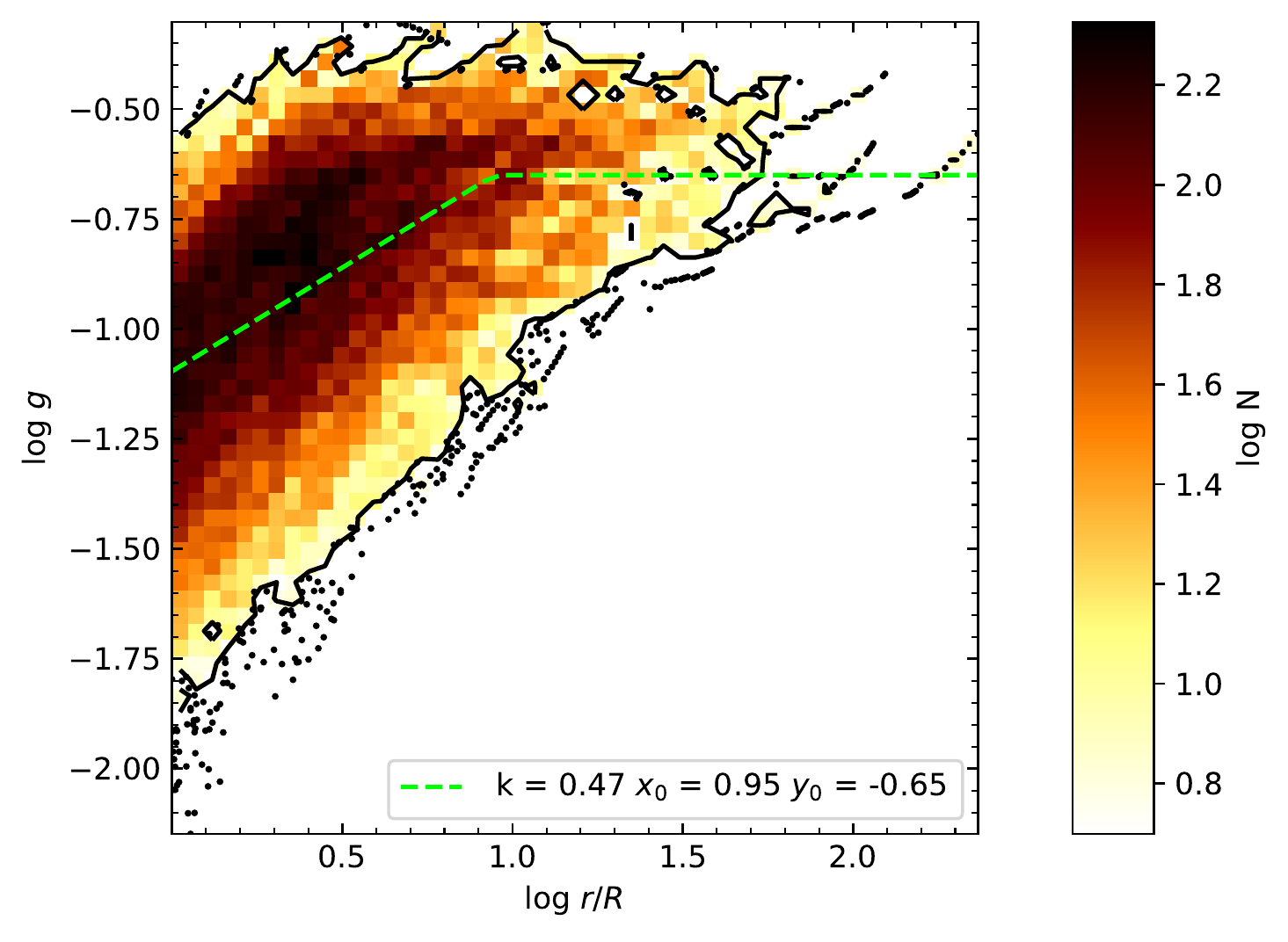}
\caption{2D distribution plot of $g$ versus $\log (r/R$). The values are determined from the smeared contours, and the color shows the number of contours in each bin (one black dot represents one single contour). The green dash line shows 
a piecewise linear fit to
this distribution.}
\label{fig: g smear}
\end{figure}

We next remove from our sample contours small enough that beam-smearing precludes an accurate estimate of $g$. To do so, for each contour we compute the ratio of the mean contour radius $r = \sqrt{A/\pi}$ to the resolution $R$ on the \textit{Herschel} maps (expressed as the effective beam FWHM, taken from Table 1 of \citealt{Pokhrel_2020}). To check when beam smearing becomes significant, we artificially blur our data: for each cloud we construct 8 column density maps smeared by Gaussian beams with sizes uniformly spaced between $R$ and $20 R$, and place contours on these smeared maps exactly as we do for the original maps. We plot the distribution of $g$ values for all contours with $\Sigma_{\rm SFR} > 0$ in \autoref{fig: g smear}. The figure clearly shows the expected effect: contours with small $r/R$ have systematically smaller values of $g$ than larger ones. To quantity this, we fit the distribution of $(\log(r/R),g)$ values with a piece-wise linear function
$g = k [\log(r/R)-x_0] + y_0$ for $\log(r/R)<x_0$, and $g=y_0$ otherwise,
where $k$, $x_0$, and $y_0$ are free parameters. The green line shows the best fit, $k=0.47$, $x_0=0.95$, and $y_0 = -0.65$, which illustrates the effect in which we are interested: the distribution of $g$ values becomes independent of resolution when $\log(r/R) \: \geq 0.95$, and drops below this threshold. We therefore remove from our analysis of the original, unsmeared maps any contour with $\log(r/R) < 0.95$. We also include only contours that contain at least one protostar, since we obviously cannot compute $\epsilon_{\rm ff}$ values for those that contain none. After this down-select we obtain 3410 contours that form the data set for further analysis.

\section{Results}
\label{sec:results}
\begin{figure}
    \includegraphics[width = \linewidth]{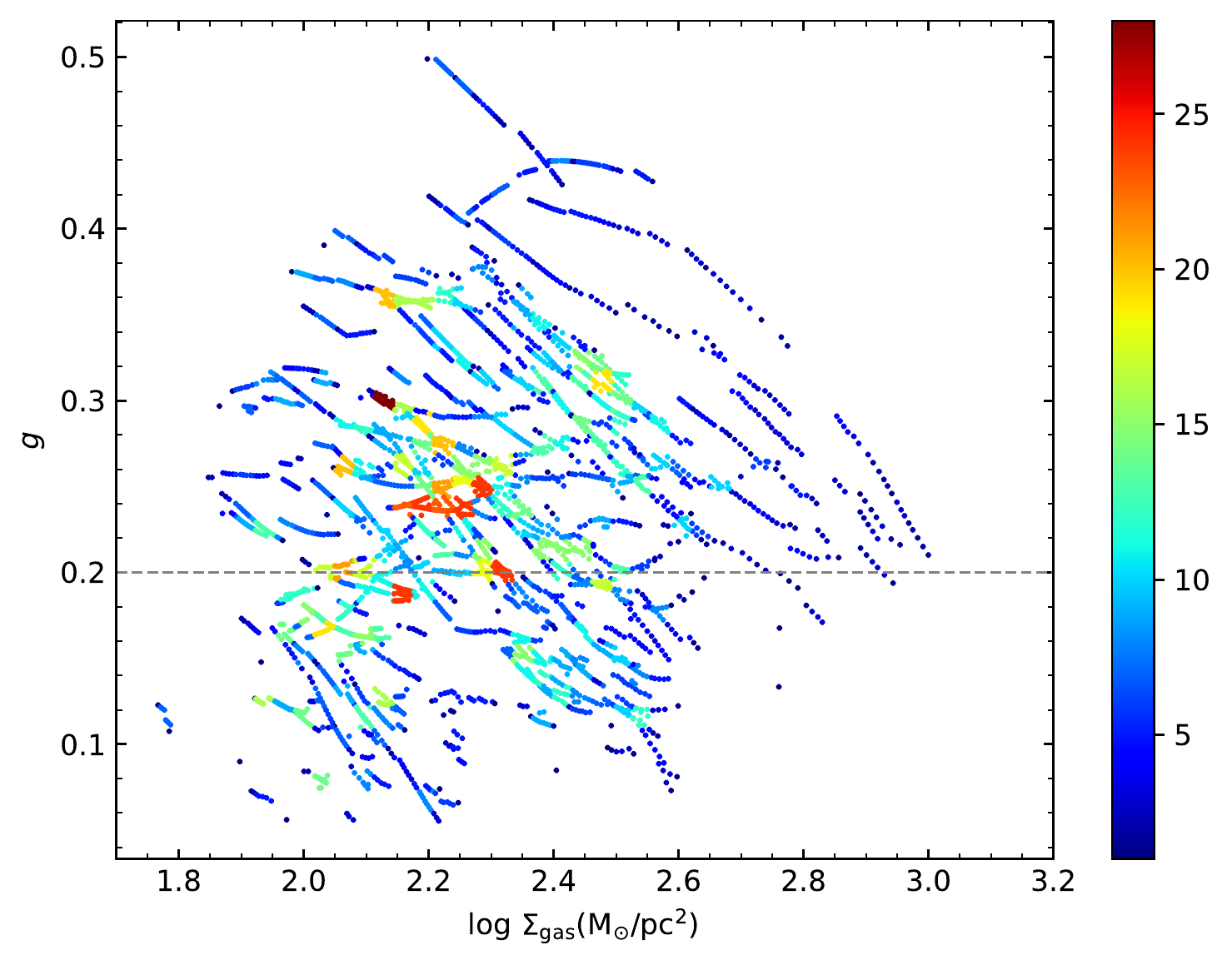}
\caption{2D scatter plot of $g$ versus log ($\Sigma_{\rm gas}$). The values are determined from the unsmeared contours, and the color shows the number of contours in the scatter circle. The grey dash line shows the $g$ = 0.2 reference line for the projection of a uniform sphere.}
\label{fig: g_sigma}
\end{figure}

With the properties determined from the unsmeared contours on different surface density levels, we first investigate how the triplet ($g$, $\epsilon_{\rm ff, sph}$, $\epsilon_{\rm ff, g}$) changes with $\Sigma_{\rm gas}$. In \autoref{fig: g_sigma} we show the scatter plot of $g$ versus log ($\Sigma_{\rm gas}$), which presents a clear trend of $g$ decreasing with $\Sigma_{\rm gas}$. It is important to note that this is \textit{not} a resolution effect, since we have removed contours small enough that lack of resolution leads to them being blurred; instead, it is a real physical trend. The median value of $g$ is $g_{\rm median} = 0.24 $, which is above 0.2, the value for a uniform-density sphere \citep{Hu_2021}. This indicates that the free-fall time of most contours will be overestimated by the spherical assumption. For a typical Gini coefficient $g = 0.24$, our estimated uncertainties on $k$ and $b$ in \autoref{eq: rho_g} translate to an increase in the dispersion of $\epsilon_{\rm ff}$ by 0.01 dex, which is negligible. For this reason, we will simply treat $k$ and $b$ as constants fixed to their central values for the remainder of this analysis.

\begin{figure}
    \begin{subfigure}{\linewidth}
        \includegraphics[width=\linewidth]{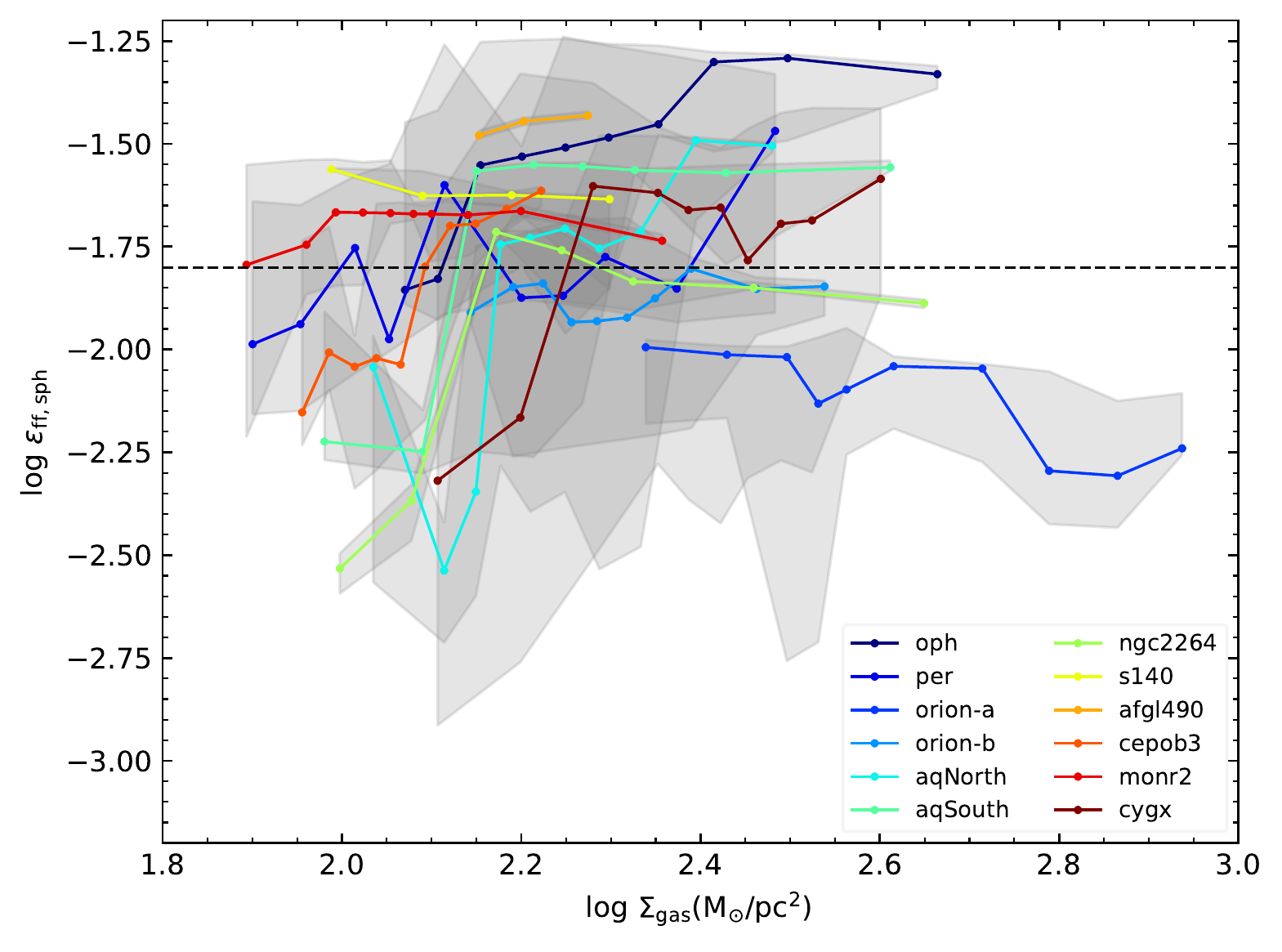} 
    \end{subfigure}
    \hfill
    \begin{subfigure}{\linewidth}
        \includegraphics[width=\linewidth]{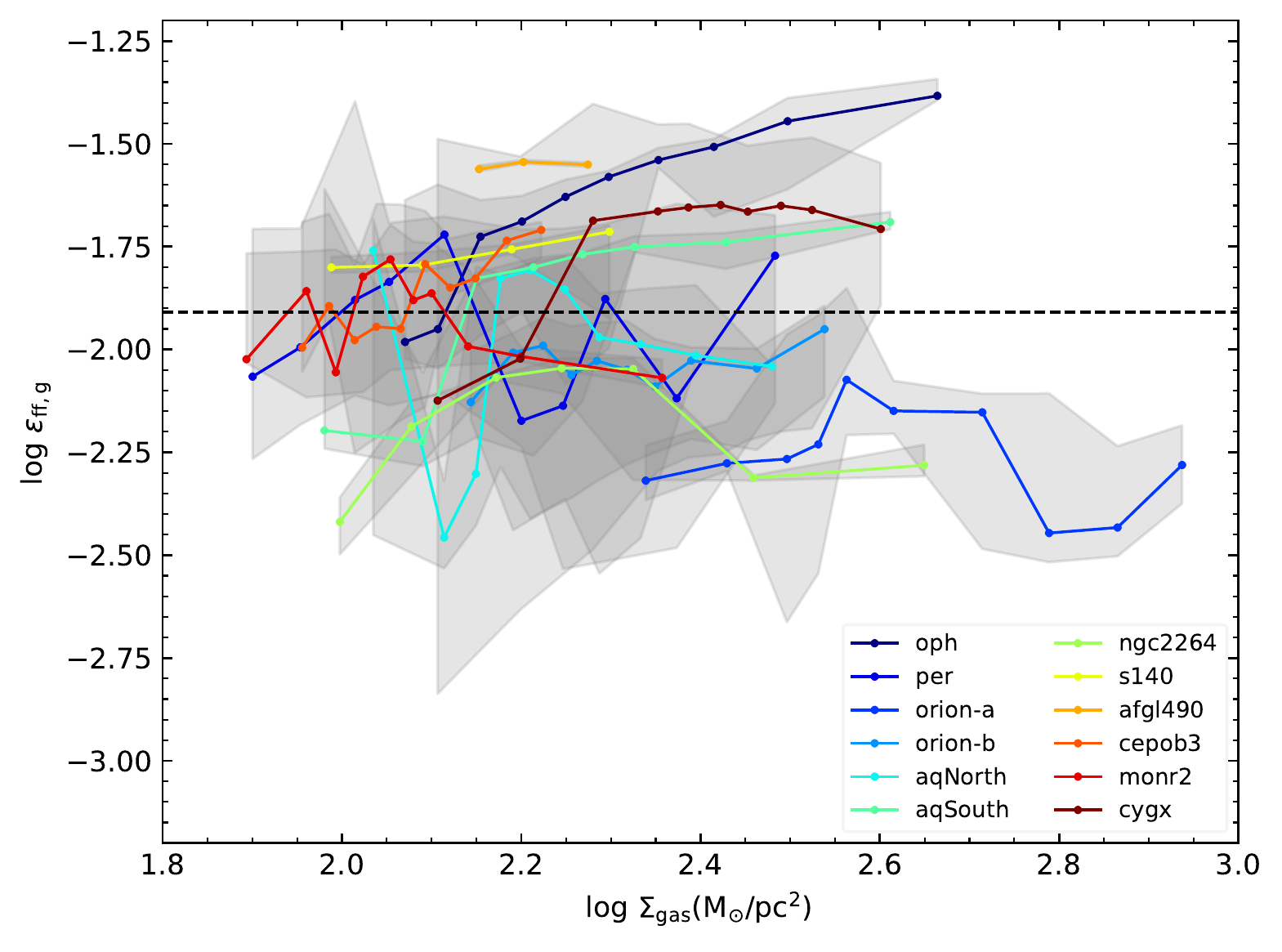} 
    \end{subfigure}
    \caption{Distributions of log $\epsilon_{\rm ff, sph}$ (top) and log $\epsilon_{\rm ff, g}$ (bottom) as a function of log $\Sigma_{\rm gas}$. For each cloud, the coloured line and grey band show the 50th percentile and 16th-84th percentile range of $\epsilon_{\rm ff}$ in a bin of $\log\Sigma_{\rm gas}$. The black dashed lines show the median values $\langle\epsilon_{\rm ff, sph}\rangle$ and $\langle\epsilon_{\rm ff, g}\rangle$ over all clouds.}
    \label{fig: EPS_SIG}
\end{figure}

To study the difference between estimates of $\epsilon_{\rm ff}$ derived using the spherical assumption and our improved method, we define two quantities: the mean star formation efficiency for an individual cloud $\langle\epsilon_{\rm ff}\rangle$, and the mean dispersion in star formation efficiency $\sigma$. We compute these as follows: for each cloud, we sort the contours by $\Sigma_{\rm gas}$ and place them in 10 bins of equal size, or fewer if that leaves a cloud with $<20$ contours per bin. (Recall that $\Sigma_{\rm gas}$ is the mean surface density inside a given contour, not the contour level itself, so two contours at the same level still generally have different $\Sigma_{\rm gas}$.) Within each bin, we denote the 16th, 50th, and 84th percentiles of $\epsilon_{\rm ff}$ (for both $\epsilon_{\rm ff,sph}$ and $\epsilon_{\rm ff,g}$) as $\epsilon_{\rm ff, 16}$, $\epsilon_{\rm ff, 50}$, and $\epsilon_{\rm ff, 84}$. We plot these quantities as a function of $\Sigma_{\rm gas}$ in \autoref{fig: EPS_SIG}. We then define the mean star formation efficiency for one cloud $\langle\epsilon_{\rm ff}\rangle$ as
\begin{equation}
    \langle\epsilon_{\rm ff}\rangle = \frac{\int_{\log\Sigma_{\rm gas, min}}^{\log\Sigma_{\rm max}} \epsilon_{\rm ff, 50}(\Sigma_{\rm gas}) \,d(\log\Sigma_{\rm gas})}{\log\Sigma_{\rm gas,max} - \log\Sigma_{\rm gas,min}},
  \label{eq: mean eps}
\end{equation}
where $\Sigma_{\rm gas,min}$ and $\Sigma_{\rm gas,max}$ are the minimum and maximum contour surface density available for a given cloud; in terms of \autoref{fig: EPS_SIG}, $\langle \epsilon_{\rm ff}\rangle$ is simply the mean value of the coloured line for each cloud. We evaluate this integral by approximating it as a finite sum over our groups. Similarly, we define the mean star formation efficiency dispersion $\sigma$ as
\begin{equation}
    \sigma = \frac{\int_{\log\Sigma_{\rm gas,min}}^{\log\Sigma_{\rm gas,max}} (\log\epsilon_{\rm ff, 84}(\Sigma_{\rm gas}) - \log\epsilon_{\rm ff, 16}(\Sigma_{\rm gas})) \,d(\log\Sigma_{\rm gas})}{\log\Sigma_{\rm gas,max} - \log\Sigma_{\rm gas,min}},
  \label{eq: eps dispersion}
\end{equation}
where we again evaluate numerically as a finite sum over our bins of $\Sigma_{\rm gas}$. In terms of \autoref{fig: EPS_SIG}, $\sigma$ is simply the mean width of the grey band that surrounds each of the coloured lines.

We report the $\langle\epsilon_{\rm ff}\rangle$ and $\sigma$ values we measure using the spherical assumption (denoted by subscript sph) and with \autoref{eq: rho_g} (subscript g) for all 12 clouds in \autoref{tab: eps_results}. After applying our model, the median value of $\log\langle\epsilon_{\rm ff}\rangle$ decreases from log$\langle\epsilon_{\rm ff,sph}\rangle = -1.80$ to $\log\langle\epsilon_{\rm ff,sph}\rangle = -1.91$. This is consistent with the prediction in \citet{Hu_2021} that use of the spherical assumption leads to a $\sim 0.13$ dex overestimate of $\epsilon_{\rm ff}$. We also measured the difference in dispersion $\Delta\sigma = \sigma_{\rm sph} - \sigma_{\rm g}$ derived using the spherical assumption versus using \autoref{eq: rho_g} for each cloud. We find that 8 of the 12 studied clouds yield positive $\Delta\sigma$, corresponding to a reduction in the dispersion; the median reduction is $\Delta\sigma_{\rm median} = 0.03$ dex. This demonstrates that our model does decrease the dispersion, but less than the $\sim 0.15$ dex found when testing the method on simulated data in \citet{Hu_2021}. This is likely due to the difference between  the simulated data and observations. \citet{Hu_2021} calibrate their method based on simulations from \citet{Cunningham_2018} that use periodic boundary conditions, so the column density maps used in the calibration are from infinitely large self-similar clouds. The observed clouds, however, are from of finite size, so, for example, they can contain large-scale density gradients that are absent in periodic boxes. This suggests that we might obtain an improved version of \autoref{eq: rho_g} by analyzing a zoom-in galactic simulation.

For all 12 clouds, we determine the standard deviations (STD) of both types of $\langle\epsilon_{\rm ff}\rangle$ values: $\text{STD}_{\rm sph} = 0.19$, and $\text{STD}_{\rm g} = 0.21$. The difference is not significant, but this is perhaps not surprising because even with the spherical assumption the dispersion of $\approx 0.2$ is very small, and may well reflect real physical differences between clouds, as we discuss below. Given the relatively small $\Delta \sigma_{\rm median}$ we obtain, it is also interesting to ask whether we could forgo individualized corrections altogether, and simply adopt the median value $g = 0.24$ for all contours. Doing so would still produce a 0.1 dex median value decrease in $\epsilon_{\rm ff}$, while leaving the dispersion unchanged. However, such an approach would miss an important subtlety: while $g = 0.24$ is the median value for all contours on all scales, the value of $g$ also changes systematically with size scale: on the largest scales of the 12 clouds we study, $g_{\rm median} = 0.35$. Properly accounting for this is crucial to obtaining the correct changes in $\epsilon_{\rm ff}$ versus $\Sigma_{\rm gas}$, and thus the correct $\Delta\sigma$ values within individual clouds. For this reason, we prefer to use individual-contour corrections when possible.

\begingroup
\setlength{\tabcolsep}{2pt} 
\begin{table}
    \centering
    \begin{tabular}{lcccccc}
    \hline
    Cloud & log$\langle\epsilon_{\rm ff,sph}\rangle$ & log$\langle\epsilon_{\rm ff,g}\rangle$ & $\sigma_{\rm sph}$ & $\sigma_{\rm g}$ & $\Delta\sigma$ & 
    $\log\Sigma_{\rm gas}$
    \\
     & & & (dex) & (dex) & (dex) & ($\rm M_{\odot} / \rm pc^2$) 
     \\
    \hline
    Ophiuchus & -1.44 & -1.57 & 0.35 & 0.25 & 0.10 & (2.05, 2.79)\\
    Perseus & -1.80 & -1.96 & 0.56 & 0.62 & -0.06 & (1.85,  2.67)\\
    Orion-A & -2.12 & - 2.27 & 0.32 & 0.31 & 0.01 & (2.24, 3.00)\\
    Orion-B & -1.87 & -2.03 & 0.38 & 0.31 & 0.07 & (2.10, 2.64)\\
    Aquila-N & -1.84 & -2.02 & 0.56 & 0.44 & 0.12 & (2.02, 2.56)\\
    Aquila-S & -1.71 & -1.86 & 0.30 & 0.23 & 0.07 & (1.92, 2.78)\\
    NGC 2264 & -1.94 & -2.19 & 0.06 & 0.11 & -0.05 & (1.97, 2.77)\\
    S140 & -1.62 & -1.77 & 0.07 & 0.07 & 0.00 & (1.95, 2.36)\\
    AFGL 490 & -1.45 & -1.55 & 0.02 & 0.01 & 0.01 & (2.14, 2.32)\\
    Cep OB3 & -1.86 & -1.86 & 0.36 & 0.30 & 0.06 & (1.92, 2.41)\\
    Mon R2 & -1.70 & -1.97 & 0.17 & 0.20 & -0.03 & (1.77, 2.48)\\
    Cygnus-X & -1.80 & -1.77 & 1.02 & 0.89 & 0.13 & (1.90, 2.78)\\
    \hline
    Median & -1.80 & -1.91 & 0.34 & 0.28 & 0.03\\
    Mean & -1.76 & -1.90 & 0.35 & 0.31 & 0.03\\
    STD & 0.19 & 0.21\\
    \hline
    \end{tabular}
    \caption{Estimates of $\langle\epsilon_{\rm ff}\rangle$ and $\sigma$ for individual clouds, using both the spherical assumption (values with subscript ``sph'') and the Gini model \autoref{eq: rho_g} (values with subscript ``g''); $\Delta\sigma = \sigma_{\rm sph} - \sigma_{\rm g}$. The column $\log\Sigma_{\rm gas}$ reports the (min, max) surface density measured for each cloud. Finally, the last three rows list the median, mean, and standard deviation values (STD) of the corresponding columns.}
    \label{tab: eps_results}
\end{table}
\endgroup

\section{Conclusions}
\label{sec:conclusions}
We use a new method proposed by \citet{Hu_2021} to combine column density maps derived from \textit{Herschel} with young stellar objects from the SESNA catalog to determine the star formation efficiency per free-fall time $\epsilon_{\rm ff}$ in 12 nearby clouds. Our method provides a more realistic estimate of the mean volume densities of clouds seen in projection, substantially reducing the error incurred by assuming that projected clouds are spherical, and allowing higher-precision estimates of $\epsilon_{\rm ff}$ than previously possible. We find that the spherical assumption leads to $\sim 0.1$ dex overestimate of $\log\langle\epsilon_{\rm ff}\rangle$, and also increases the estimated intra-cloud dispersion in $\log\langle\epsilon_{\rm ff}\rangle$ by $\sim 0.03$ dex on average. With our new method, we find that our sample of 12 clouds has a median star formation efficiency per free-fall time $\log\langle\epsilon_{\rm ff}\rangle = -1.9$, and the median spread in $\log\langle\epsilon_{\rm ff}\rangle = 0.28$ dex within a single cloud. The inter-cloud dispersion in $\log\langle\epsilon_{\rm ff}\rangle$ is nearly identical, at 0.21 dex, and this value is, within the uncertainties, unaffected by the use of the \citet{Hu_2021} model for the gas density. This strongly suggests that the intra-cloud dispersion we are measuring reflects a real variation in cloud properties, not an observational error.

Our results confirm the existence of a universal $\epsilon_{\rm ff} \sim 0.01$ value, and, importantly, let us identify a real $\approx 0.2$ dex spread from cloud to cloud with 3D cloud geometry considered for the first time. As discussed in \citet{Pokhrel_2021}, such a small spread is in tension with models where star formation is regulated mainly by galactic-scale processes, but individual molecular clouds undergo rapid collapse. These models predict a much larger dispersion. Conversely, however, our measured spread in $\epsilon_{\rm ff}$ can be used to evaluate the spread in parameters that enter models for cloud-scale regulation of star formation, which do predict dispersions comparable in size to the observed one. For example, in the turbulence regulated star formation of \citet{Krumholz_2005}, a $\sim 0.2$ dex spread in $\epsilon_{\rm ff}$ could naturally be explained by a $\sigma_{\alpha} \sim 0.3$ dex spread in cloud virial parameters (or a  $\sigma_{\mathcal{M}}\sim 0.7$ dex spread in Mach number), while in the similar model of \citet{Hennebelle_2011} the required dispersion is $\sigma_{\alpha} \sim 0.8$ dex ($\sigma_{\mathit{M}} \sim 0.6$ dex), and for the model of \citet{Padoan_2012} would require $\sigma_{\alpha} \sim 0.5$ dex. For comparison, \citet{Lee_2016} study 195 star forming giant molecular clouds and find a scatter of $0.32$ dex in the virial parameter. Thus observed clouds have approximately the level of dispersion in virial parameter required to reproduce the spread we see in $\epsilon_{\rm ff}$. In future work, we can use the same technique of high-precision estimates of $\epsilon_{\rm ff}$ deployed here to search not just for the dispersion in $\epsilon_{\rm ff}$, but to look for systematic variations with virial parameter or other cloud properties, thereby opening up a new method for testing theories of star formation.
\section*{Acknowledgements}

MRK acknowledges funding from Australian Research Council awards DP190101258 and FT180100375. RP and RAG acknowledge support from NASA ADAP awards NNX15AF05G, 80NSSC18K1564 and NNX17AF24G. RP acknowledges funding from NASA ADAP award 80NSSC18K1564, and RAG acknowledges funding from NASA ADAP awards NNX11AD14G and NNX13AF08G.  We further acknowledge high-performance computing resources provided by the Australian National Computational Infrastructure (grants~jh2 and ek9) through the National and ANU Computational Merit Allocation Schemes, and by the Leibniz Rechenzentrum and the Gauss Centre for Supercomputing (grant~pr32lo).

This research has made use of data from the \href{http://gouldbelt-herschel.cea.fr}{Herschel Gould Belt survey (HGBS) project}, a Herschel Key Programme jointly carried out by SPIRE Specialist Astronomy Group 3 (SAG 3), scientists of several institutes in the PACS Consortium (CEA Saclay, INAF-IFSI Rome and INAF-Arcetri, KU Leuven, MPIA Heidelberg), and scientists of the Herschel Science Center (HSC).

\section*{Data Availability}

The data underlying this article will be shared upon reasonable request to the corresponding author.



\bibliographystyle{mnras}
\bibliography{Hu2021b} 




\bsp	
\label{lastpage}
\end{document}